\documentclass[usenatbib,usegraphicx]{mn2e}
\usepackage{amssymb}
\usepackage[fleqn]{amsmath}
\usepackage{charter}
\usepackage{mathdesign}
\usepackage{url}
\usepackage{bookmark}
\providecommand{\eprint}[1]{\href{http://arxiv.org/abs/#1}{#1}}
\providecommand{\adsurl}[1]{\href{#1}{ADS}}
 
\usepackage{natbib}

\newcommand{\nat}   {Nature}

\newcommand{\memsai} {Mem.~Soc.~Astron.~Italiana}
\newcommand{\na}   {NewA}

\usepackage{array}
\usepackage{graphicx}
\usepackage{subfigure}
\usepackage[usenames,dvipsnames]{xcolor}
\usepackage{color}

\graphicspath{{figures/}}

\pagerange{\pageref{firstpage}--\pageref{lastpage}} \pubyear{2011}

\def\LaTeX{L\kern-.36em\raise.3ex\hbox{a}\kern-.15em
T\kern-.1667em\lower.7ex\hbox{E}\kern-.125emX}

\newcommand{\B}{\begin{eqnarray}}
\newcommand{\E}{\end{eqnarray}}

\title[Recollimation Boundary Layers in Relativistic Jets]{Recollimation Boundary Layers in Relativistic Jets}

\author[Kohler et al.]
{Susanna Kohler$^{1, 2 \ast}$,
Mitchell C. Begelman$^{1, 2 \star}$,
Kris Beckwith$^{1, 3 \dagger}$
\\$^1$ JILA, University of Colorado and National Institute of Standards and Technology, Boulder, CO 80309-0440, USA 
\\$^2$ Department of Astrophysical and Planetary Sciences, University of Colorado, Boulder, CO 80309-0391, USA
\\$^3$ Tech-X Corporation, 5621 Arapahoe Ave. Suite A, Boulder, CO 80303, USA
\\Email: $^\ast$ kohlers@colorado.edu,
$^\star$ mitch@jila.colorado.edu,
$^\dagger$ kris.beckwith@jila.colorado.edu}

\begin{document}

\label{firstpage}

\maketitle

\begin{abstract} 
We study the collimation of relativistic hydrodynamic jets by the pressure of an ambient medium in the limit where the jet interior has lost causal contact with its surroundings.  For a jet with an ultrarelativistic equation of state and external pressure that decreases as a power of spherical radius, $p \propto r^{-\eta}$, the jet interior will lose causal contact when  $\eta>2$.  However, the outer layers of the jet gradually collimate toward the jet axis as long as  $\eta < 4$, leading to the formation of a shocked boundary layer. Assuming that pressure-matching across the shock front determines the shape of the shock, we study the resulting structure of the jet in two ways: first by assuming that the pressure remains constant across the shocked boundary layer and looking for solutions to the shock jump equations, and then by constructing self-similar boundary-layer solutions that allow for a pressure gradient across the shocked layer. We demonstrate that the constant-pressure solutions can be characterized by four initial parameters that determine the jet shape and whether the shock closes to the axis. We show that self-similar solutions for the boundary layer can be constructed that exhibit a monotonic decrease in pressure across the boundary layer from the contact discontinuity to the shock front, and that the addition of this pressure gradient in our initial model generally causes the shock front to move outwards, creating a thinner boundary layer and decreasing the tendency of the shock to close. We discuss trends based on the value of the pressure power-law index $\eta$. 
\end{abstract}



\section{Introduction}

The idea that AGN outflows are highly collimated is supported by observations (e.g. \citealp{MB84}, \citealp{Jorstad05}), implying that confinement must occur. The cause of this confinement, which may occur at distances of just a few tens of Schwartzschild radii from the central black hole (\citealp{JunorBiretta99}), is however not yet well-understood. It is generally accepted that some collimating agent is necessary, but this still allows for a variety of possibilities, both external (e.g. pressure confinement via an ambient static medium \citealp{Eichler82}, \citealp{Komissarov97}, inertial confinement via a slow outflow or wind surrounding the jet \citealp{KomissarovRamPressure94,BL07}) and internal (magnetic confinement via hoop stress due to the toroidal magnetic field component \citealp{Benford78}, \citealp{MB84}).

Pressure confinement is of particular interest because accretion disk winds surrounding an AGN provide an ideal external medium for interaction with the jet. Moreover, another collimating agent such as magnetic hoop stress cannot function alone; without an ambient medium to confine the globally-expansive magnetic field, collimation will not occur (\citealp{MB95,KomBarkVla07,KomVlaKon09,Komissarov11}). Thus pressure confinement may be relevant both on its own and in conjunction with other~processes.

In this context, we note that numerical simulations that study confinement and acceleration of the flow by MHD processes treat the action of the external medium as an applied boundary condition (e.g. \citealp{KomBarkVla07,KomVlaKon09,Komissarov11}). In this work, we study the detailed physics of jet collimation by the external medium as a first step towards a complete treatment of both the action of the external medium and magnetic effects in collimating and accelerating relativistic jets.

The location and mechanism of jet collimation is interesting because it provides insight into energy dissipation within the jet. In a steady jet, there are two main sites where energy dissipation is likely to occur: at the jet spine, as a result of kink instabilities driven by the toroidal magnetic field (\citealp{MB98}, \citealp{Eichler93}), and at the interface between the jet and its environment, as a result of shear instabilities (\citealp{Micono00,Perucho10,Barkov11}) or collimation shocks, as we discuss here.

In the case of a jet with an opening half-angle of $\theta$ and a bulk Lorentz factor of $\Gamma$, the outer edge of the jet remains in causal contact with the jet center only if the opening angle obeys $\Gamma^{-1} > \sin \theta \approx \theta$. In this work, we assume a relativistic equation of state and examine the case where causal contact between the jet's spine and edge has been lost. Thus we model the jet as having a transverse structure consisting of two components: an inner region in which the flow is undergoing free expansion as it accelerates, and an outer shocked boundary layer region that results from the loss of causal contact within the jet. The geometrical shape that the jet assumes as it propagates is a direct result of its response to the collimating forces exerted by the ambient medium, and calculating that shape for the case of pressure confinement will be a major focus of this paper. 

Our goal is to determine the basic jet structure and geometry under a simple set of collimation and acceleration assumptions, providing a model for the jet's steady-state configuration. This will allow more realistic initial conditions for simulations of instabilities and turbulence, which will in turn provide a model for AGN jets to which we can compare observational signatures.

While we specifically reference AGN jets in this work, our results can easily be extended to other relativistic outflows, such as a gamma-ray burst in the collapsar model, collimated by the stellar envelope it breaks through, as discussed in \citealp{BL07} (hereafter BL07).

In this paper we use the shock conditions for a hydrodynamic, relativistic jet to derive the basic structure of the jet. This problem has been previously studied in the case of a "cold" (inertia-dominated) jet (\citealp{Komissarov97}, \citealp{KN09}), but we now focus on the collimation behavior of a "hot" (pressure-dominated) jet.

We initially follow a similar approach to that of BL07, but deviate from its methods in our treatment of entropy distribution within the jet. \citealp{BL09} (BL09) performs a similar analytical inspection in the limit of small angles; we generalize this to all angles. 

In \S 2 we find solutions for the jet shape using the Kompaneets approximation. In  \S 3 we examine self-similar solutions for the boundary layer when a pressure gradient is allowed to form across the layer, and then revise our solutions from \S 2 to include this pressure gradient. In \S 4 we conclude, summarizing the results and discussing future work.


\section{Kompaneets Approximation}

We consider a cylindrically symmetric, ultrarelativistic jet injected into an ambient medium that has a power-law pressure profile. We wrap the physics of the shocked ambient medium into this external pressure profile and focus on the structure of the jet itself. In this stage of the treatment, we ignore magnetic fields and assume that the external pressure due to the ambient medium creates the sole collimating force on the jet. 

We define $R$ and $z$ as dimensionless parameters in cylindrical coordinates describing the radial and axial distances as scaled by $z_0$, the height at which the jet initially encounters the external medium. The jet is injected with an initial opening half-angle of $\theta_0$ and impacts the wall of the ambient medium at the point thus denoted as $(R_0 = \tan \theta_0,  z_0 = 1)$. We assume the jet is injected from a point source with steady flow, and streamlines are conical and characterized by the angle $\theta_j$.

We further suppose that the interior of the jet is undergoing free expansion. As relativistic adiabatic expansion obeys $pV^{4/3} \sim const$, it therefore exhibits a corresponding pressure profile of $p_j \propto r^{-4}$, where $r$ is the spherical radius defined as $r \propto (R^2 + z^2)^{1/2}$ in cylindrical coordinates. The gradual acceleration via adiabatic expansion of the jet interior is governed by the relativistic Bernoulli equation, which describes the conversion of thermal to kinetic energy as $p \propto \Gamma^{-4}$ along streamlines (see, e.g., \citealp{Landau59}).

\begin{figure}
\center
\includegraphics[width=3.0in]{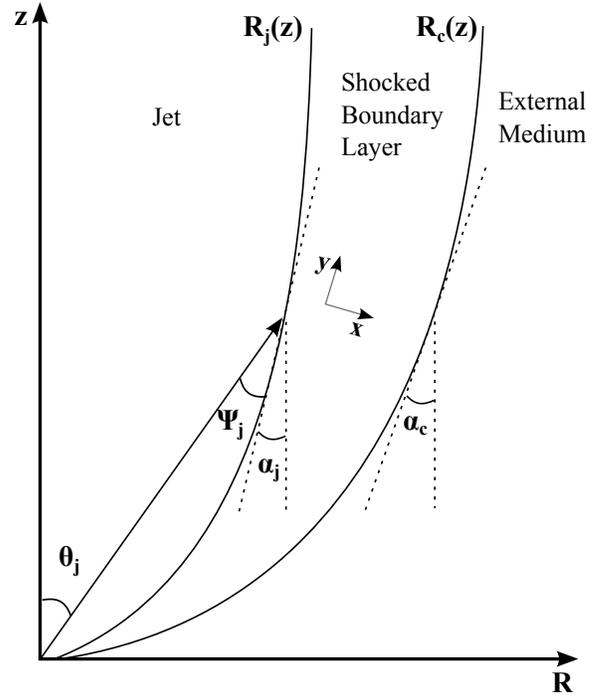}
\caption{Diagram indicating the angles and orientation of the axes relative to the inner shock wall. $\theta_j$ is the angle the jet streamline makes with the $z$-axis. $\Psi_j$ is the angle it makes with the shock normal, and $\alpha_j$ is the angle between the shock normal and the vertical.}
\label{fig:jetdiagram}
\end{figure}

Where the jet impacts the ambient medium with a supersonic normal velocity, a shocked layer will form, as indicated schematically in Figure \ref{fig:jetdiagram}. The layer is bounded on the inside by a shock front, and on the outside by a contact discontinuity. There is no mass flux across the contact discontinuity, and the pressure must be matched on either side of it. Adopting a pressure profile for the ambient medium of $p_s \propto r^{-\eta}$, for a parameter $\eta$, this fixes the pressure external to the jet and immediately inside the contact discontinuity.

We now adopt the Kompaneets approximation (\citealp{Kompaneets60}; see e.g. BL07 and \citealp{Komissarov97}), treating the pressure as a function only of $z$ within the shocked layer. Thus the pressure profile of the external medium extends across the shocked layer with no pressure gradient in the axial direction, greatly simplifying the problem.

In this paper, we focus particularly on the less-explored case of an external pressure profile with $2 < \eta < 4$ (see e.g. BL07 for an example of treatment of this regime). An $\eta < 2$ implies that if the jet begins in causal contact it will remain in causal contact, suggesting that a shocked boundary layer would not form. For $\eta = 4$ the ambient pressure profile is equivalent to that of free expansion, matching the pressure within the jet, and an $\eta > 4$ would result in a rarefaction at the jet boundary (\citealp{MB84}). Thus the regime between these two values is a logical place to examine.

Physically, this pressure profile range could describe multiple scenarios for the confining medium. The ram pressure of a head-on wind decreases as $p \propto r^{-2}$, and any obliquity would serve to steepen that pressure profile (\citealp{Eichler82}). The range is similarly relevant if the confining medium were an accretion flow such as Bondi accretion ($p \propto r^{-5/2}$), and it would not be an unrealistic range for a disk corona, or a stellar envelope in a GRB collapsar model.

The flow parameters across the inside boundary of the shocked layer are governed by the relativistic oblique shock-jump conditions. These establish conservation of mass, momentum, energy, and tangential velocity:
	\B
	n_j \Gamma_j \beta_{j,x} &=& n_s \Gamma_s \beta_{s,x}  \label{jump_j} \\
	w_j \Gamma_j^2 \beta_{j,x} &=& w_s \Gamma_s^2 \beta_{s,x}  \label{jump_s} \\
	w_j \Gamma_j^2 \beta_{j,x}^2 + p_j &=& w_s \Gamma_s^2 \beta_{s,x}^2 + p_s  \label{jump_3} \\
	\beta_{j,y} &=& \beta_{s,y}    \label{jump_4}
	\E 
where the $x$-direction is chosen perpendicular to the shock front and the $y$-direction is tangential (see Figure \ref{fig:jetdiagram}). The subscript $j$ denotes quantities within the freely-expanding inner jet region, and the subscript $s$ denotes quantities within the shocked boundary layer. Here $\beta = v/c$, and $w \equiv \epsilon + p$ where $\epsilon$ is the total proper energy density, given by $\epsilon = \rho + 3 p$ with $\rho$ defined as the proper rest mass density. Considering the case of an ultrarelativistic gas in the regime where the jet is still accelerating, we assume that $\rho \ll p$ and the equation of state is $p=\epsilon/3$, such that  $w \approx 4p$. 

Examining the geometry of the problem, we recognize that the angle of impact of the streamline with the shock front, $\sin \Psi_j$, can be represented in terms of $\theta_j$ and $\alpha_j$, respectively the angle the streamline makes with the $z$-axis and the angle that the shock tangent makes with the $z$-axis. By noting that $\tan \alpha_j = {dR_j}/{dz}$, where $R_j$ is the shape of the shock front, and assuming conical streamlines such that $\tan \theta_j = {R_j}/{z}$, we can combine the geometry of the problem with the general shock jump conditions to obtain a differential equation governing the shape of the inner shock wall,
	\B
	\dfrac{\left ( R_j - z \frac{dR_j}{dz} \right )^2}{\left ( R_j^2 + z^2 \right ) \left ( 1 + \left ( \frac{dR_j}{dz} \right )^2 \right )}
	&=& \dfrac{1}{8 \Gamma_j^2} \left (3 \dfrac{p_s}{p_j} + 1 \right ).   \label{finalde}
	\E
This can be solved analytically for ${dR_j}/{dz}$ and numerically integrated to find $R_j(z)$.

To solve for the shape of the contact discontinuity separating the shocked layer from the ambient medium, denoted as $R_c$, we assume energy-momentum conservation through a volume of the shocked layer, as described in BL07. As in BL07, we assume that the flow parameters within the shock depend only on the vertical distance $z$ in order to make the problem analytically tractable. Making the further assumption that $|\boldsymbol{\beta}_j| \approx 1$, we obtain a differential equation similar to that in BL07 governing the contact discontinuity:
	\B
	\dfrac{d}{dz} \left ( p_s \Gamma_s^2 \beta_{s, z}( R_c^2 - R_j^2 ) \right ) &=& 2 p_j \Gamma_j^2 R_j \dfrac{\sin \Psi_j}{\cos \alpha_j}. 	\label{outerwall}
	\E 
	
We insert into this the known expressions for $p_s$, $p_j$ and $\Gamma_j$ from the previous paragraphs. While a shock exists, $\Gamma_s$ and $\beta_{s,z}$ are obtained from the original shock jump conditions; if the shock closes to the axis then  $\Gamma_s$ and $\beta_{s,z}$ are obtained by assuming adiabatic expansion within the fully-shocked jet. We then numerically integrate  Eq \eqref{outerwall} simultaneously with the differential equation for the inner shock wall, solving for both $R_j(z)$ and $R_c(z)$. Thus we obtain the shapes of the shock front and the contact discontinuity in terms of the initial pressure ratio ${p_{s,0}}/{p_{j,0}}$, the initial opening angle $\theta_0$, the initial Lorentz factor $\Gamma_{j,0}$, and the pressure power-law index~$\eta$.

\begin{figure}
\center
\subfigure[$\eta~=~7/3$]{
	\includegraphics [width=1.5in] {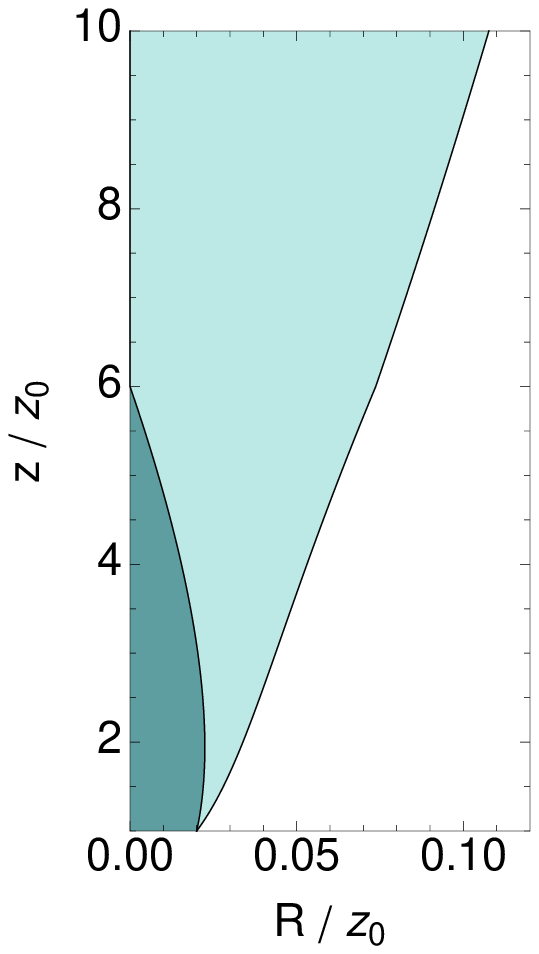}
	\label{fig:fig2_1}
	}
\subfigure[$\eta~=~11/3$]{
	\includegraphics [width=1.5in] {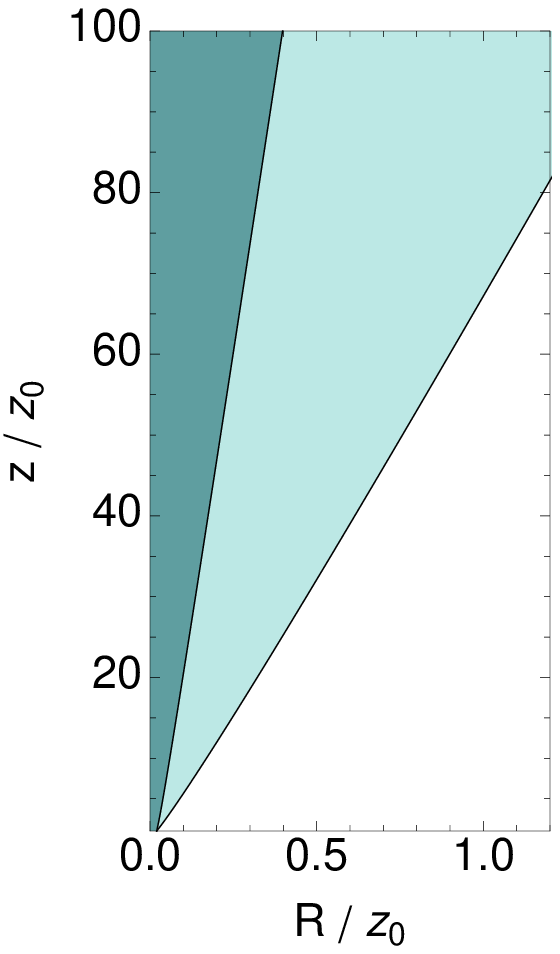}
	\label{fig:fig2_2}
	}
\caption{Two examples of the solutions for the shock front and associated contact discontinuity. In the first plot, $\eta = 7/3$; in the second $\eta=11/3$. For both plots the remaining parameters are held constant at ${p_{s,0}}/{p_{j,0}}~=~1$, $\Gamma_{j,0}=50$, and $\theta_0=1/50$.}
\label{fig: fig2final}
\end{figure}

Examining the effects of varying these four parameters, we see that under certain conditions the external pressure can drive the shock front back to the jet axis, resulting in a fully-shocked jet. The jet is more likely to close when ${p_{s,0}}/{p_{j,0}}$ is large and $\eta$ is small. Initial under- or over-pressurization of the jet strongly affects whether or not it closes, but for physical scenarios we would expect that the pressure is approximately balanced where the jet first impacts the wall, ${p_{s,0}}/{p_{j,0}} = 1$. Increasing $\eta$ can change whether or not the shock will reach the axis and, for cases where the jet does close, drives the point at which this occurs down the $z$-axis, further from the source. Two examples are shown in Figure \ref{fig: fig2final}: one in which the shock converges to the axis and one in which it doesn't. This demonstrates the effect that $\eta$ can have when all other parameters are held constant.

\begin{figure}
\center
\includegraphics [width=3.3in] {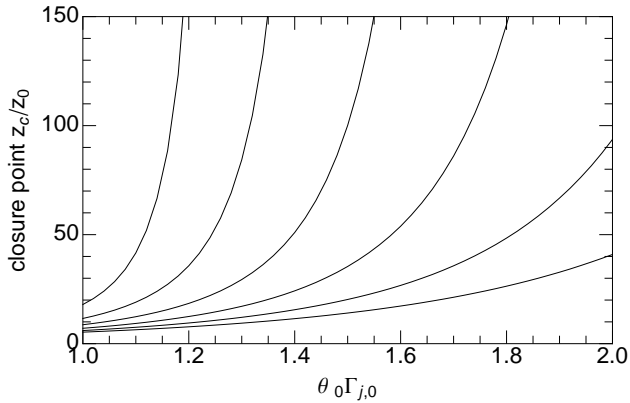}
\caption{Dependence of the jet closure point on $\theta_0 \Gamma_{j,0}$. The different curves correspond to various values of $\eta$, ranging from $\eta = 13/6$ on the right to $\eta = 3$ on the left, in increments of $1/6$. Here again, ${p_{s,0}}/{p_{j,0}}~=~1$. The closure point does not depend on the value of $\Gamma_{j,0}$ individually.}
\label{fig:tg}
\end{figure}

Whether or not a jet closes and, if it does, the value of $z$ at the point where the shock meets the axis is dependent upon the value of the product $\theta_0  \Gamma_{j,0}$, rather than on $\theta_0$ or $\Gamma_{j,0}$ individually. Assuming that the initial pressure ratio and $\eta$ are held constant, we find that all jets with the same value of $\theta_0 \Gamma_{j,0}$ close at the same point on the $z$-axis. Increasing $\theta_0 \Gamma_{j,0}$ causes the jet to close further down the axis from the source, until the point where it no longer closes. This dependence of the closure point on $\theta_0 \Gamma_{j,0}$ is illustrated in Figure \ref{fig:tg} for $2<\eta \le 3$, assuming ${p_{s,0}}/{p_{j,0}} = 1$. For $\eta > 3$, $\theta_0 \Gamma_{j,0} < 1$ is required to produce a jet that closes.

We find that the shape of the contact discontinuity is initially dependent upon the values of the four parameters ${p_{s,0}}/{p_{j,0}}$, $\theta_0$, $\Gamma_{j,0}$, and $\eta$. For large $z$, however, the contact discontinuity asymptotes to one of two shapes: if the shock closes to the axis, then the flow is governed by free expansion beyond that point and the contact discontinuity takes the shape $R_c \propto z^{\eta/4}$, as found in previous works such as BL07, BL09 and \citealp{Levinson00}. If the shock does not close to the axis, however, then the contact discontinuity takes the shape $R_c \propto z$, in contrast to these previous works.

The discrepancy between our work and previous studies appears to arise as a result of entropy treatment: in our work, we assume that as long as the shock has not yet closed to the axis, the continued addition of material into the boundary layer ensures that the boundary layer cannot have constant entropy throughout, and thus the layer is not governed by adiabatic expansion. We instead solve for all quantities directly from the shock jump equations, without making assumptions about the behavior of material within the boundary layer.

This difference in the behavior while the jet remains open is crucial, since this model for the jet can only be accurately applied in the regime in which it remains open. After the jet has closed, complex effects such as rarefaction waves or oblique shock reflections (e.g. \citealp{Gomez95}) will likely arise, and this simple model no longer adequately describes the jet's behavior. 

 The conical asymptote that we find in this approximation will be an important factor in our ability to refine this model, as shown in \S 3.


\section{Self-Similar Treatment of the Boundary Layer}

The Kompaneets assumption of constant pressure across the shocked layer is inconsistent with basic intuition: in a realistic large-scale jet, one would expect that the curvature of the streamlines as the jet collimates would go hand in hand with a force inwards along the radius of curvature. Because of the special-relativistic length contraction along the direction of motion, the curvature appears to the fluid to be more extreme than it is in the lab frame by a factor of $\Gamma$, resulting in a sizable centripetal force for even slight curvature. Due to this centripetal force, a pressure gradient should then form across the boundary layer, with higher pressure at the outer wall of the jet and lower pressure at the shock front. 

With this in mind, we now refine the model in \S 2 by including effects of a transverse pressure gradient within the boundary layer. Even assuming a steady state and axial symmetry, this problem intrinsically involves the solution of partial differential equations in two dimensions, so we simplify the problem by seeking self-similar solutions. In order to do so, we treat the streamlines as being very nearly conical (an assumption justified by the results of \S 2) and calculate their deviation from conical as a function of position within the boundary layer, thus examining the effects of a pressure gradient across the boundary layer.


\subsection{Self-Similar Solutions}

We assume that the opening angle of the jet is much greater than $1/\Gamma$, such that causal contact has been lost. We further suppose that the boundary layer that forms has a thickness that is of order $\Delta \theta \sim 1/\Gamma$, ensuring that the boundary layer is very thin compared to the width of the jet. The radius of curvature of the jet is then much larger than the width of the boundary layer, allowing us to treat the curvature as a small effect. 

Given the assumption of nearly conical streamlines, we can in this case treat the entropy and the Bernoulli constant as being the same on all streamlines, implying potential flow. These assumptions are physically realizable if the majority of the material within the shock enters at approximately the same point near the base of the jet; we will later demonstrate that this assumption is self-consistent here.

The flow within the boundary layer is governed by the energy equation, momentum conservation along streamlines, and mass conservation:
	\B
	\rho &=& A p^{3/4}  \label{ssenergy} \\
	\dfrac{\Gamma w}{p^{3/4}} &=& B  \label{ssmomentum} \\
	\boldsymbol{\nabla} \cdot (\rho \boldsymbol{\beta} \Gamma) &=& 0  \label{ssmass}
	\E 
where $A$ and $B$ are constants. We henceforth treat the problem in the spherical polar coordinates $r, \theta$ and $\phi$ for convenience in describing a jet with approximately radial streamlines. As the maximum transverse speed that can be achieved without a shock forming is of order ${1}/{\Gamma}$, we can therefore assume that $\beta_\theta$ is of this order, and $\beta_r$ is of order one. Adopting this characteristic scale, we state that $\frac{\partial}{\partial \theta} \sim \Gamma \frac{\partial}{\partial r} $. Writing out $\beta_r$ and employing the fact that $\beta_\theta^2 + \Gamma^{-2} \ll 1$, we have $\beta_r \approx 1- \frac{1}{2} ( \beta_\theta^2 + \Gamma^{-2})$. Using this, we then combine the flow equations, retaining terms only to lowest order. 

If we now assume that the external pressure is a power law in spherical radius $r$, $p_e \propto r^{-\eta}$, then we can find self-similar solutions for the flow structure near the wall of the jet in the specific case of a pressure-dominated jet, or near enough to the jet source that the Lorentz factor is much less than its asymptotic value ($\Gamma \ll \Gamma_\infty$). In this limit, the equations reduce~to a pair of coupled partial differential equations for the Lorentz factor $\Gamma$ and the transverse velocity $\beta_\theta$ within the boundary layer:
	\B
	\dfrac{1}{r} \dfrac{\partial}{\partial r} \left (\dfrac{r^2}{\Gamma^2} \right) +  \dfrac{\partial}{\partial \theta} \left (\dfrac{\beta_\theta}{\Gamma^2} \right ) &=& 0      \label{sspresslim1}   \\
	\dfrac{\partial}{\partial r}(r \beta_\theta) + \beta_\theta \dfrac{\partial \beta_\theta}{\partial \theta} + \dfrac{1}{2} \dfrac{\partial}{\partial \theta} \left (\dfrac{1}{\Gamma^2} \right ) &=& 0.      \label{sspresslim2}  
	\E

We now seek self-similar solutions of the form
	\B
	\dfrac{1}{\Gamma} = r^{-\eta/4} g(\xi), \phantom{m} p = r^{- \eta} g^4(\xi),  \phantom{m} \beta_\theta = r^{-\eta/4} h(\xi), \label{ssgammaeq}
	\E
where we have chosen the constant in the Bernoulli equation to be unity (i.e. $p \Gamma^4 = 1$) for simplicity. In these solutions $g$ and $h$ are functions of a similarity variable $\xi$ that describes the distance from the contact discontinuity, normalized by the expected scale of the boundary layer, $\xi \propto (\theta_c - \theta)/{\Delta \theta}$. The angular thickness of the boundary layer is expected to scale as $\Delta \theta = {1}/{\Gamma_c}$, such that $\xi \propto r^{\eta/4} (\theta_c - \theta)$, where $\theta_c = \theta_c(r)$ is the location of the contact discontinuity.

The fact that the streamlines at $\theta_c$ must be parallel to the contact discontinuity, requiring that $\beta_\theta(\theta_c) = r d\theta_c / dr$, yields the further constraint that
	\B
	\dfrac{d\theta_c}{dr} &=& h(0) r^{-(1+\eta/4)}.
	\E
	
Choosing the proportionality constant such that $\xi$ is defined as
	\B
	\xi = - \dfrac{1}{h(0)} r^{\eta/4} (\theta_c - \theta)
	\E
absorbs the boundary condition into the similarity variable and ensures collimating solutions (such that $h(0)<0$).
	
The boundary condition $g(0) = 1$ is enforced so that the pressure is matched at the contact discontinuity, but $h(0)$ is allowed to range. Assuming solutions of this form, we define 
	\B
	w(\xi)=\frac{g(\xi)}{g(0)}, \phantom{i} q(\xi)=\frac{h(\xi)}{h(0)}  \text{  and  }   \mu = \left(\frac{g(0)}{h(0)}\right)^2 
	\E
 with $w(0) = q(0) =1$. Further defining $x = (1-\eta \xi/4)$, we obtain a set of coupled linear ordinary differential equations for $w(x)$ and $q(x)$ that can be cast to reflect the existence of a critical point: 
	\B
	(\mu w^2)' &=& \dfrac{\frac{8}{\eta}(1-\frac{\eta}{4})(2x - q) \mu w^2}{\mu w^2 - 2 (x -q) ^2}     \label{critpt1} \\
	q' &=& \dfrac{\frac{8}{\eta}(1-\frac{\eta}{4}) \left( \mu w^2+ q(x - q) \right)}{\mu w^2 - 2 (x -q) ^2}.	\label{critpt2}
	\E

The critical point occurs where the denominator of these equations goes to zero, forcing the numerators to also go to zero at this point in order to prevent $q'$ and $(\mu w^2)'$ from diverging.

The critical point can be physically understood as a type of sonic point, in analogy with the critical points discussed in \citealp{Blandford82} (BP82). Setting the denominator in Eqs \eqref{critpt1} and \eqref{critpt2} equal to zero yields the constraint that $\Gamma^2 \beta_\xi^2 = 1/2$ at the critical point, where $\beta_{\xi} = \boldsymbol{\beta} \cdot \boldsymbol{\nabla} \xi / \lvert \boldsymbol{\nabla} \xi \rvert$. This constraint is mathematically equivalent to two conditions: that the speed of sound waves measured in the lab frame in the direction of $\boldsymbol{\nabla} \xi$ must vanish (analogous to BP82's condition that the sound speed normal to the similarity surfaces is equal and opposite to the component of flow speed in the same direction), and that the sound speed along the similarity surfaces is equal to the fluid speed and carries no signals in this direction (analogous to BP82's condition that the wave signals are normal to the similarity surfaces).

The physical solutions for $q$ and $w$ are those that pass through the critical point; the solutions that don't pass through the critical point display unphysical behavior, such as crashing and becoming double-valued.

Requiring finite $q'$ and $(\mu w^2)'$ at the critical point results in three possible sets of relations among $q, w, \mu$ and $x$ specifically at the critical point. Each one of these sets of relations at the critical point applies uniquely to one of the three regions $\eta < 8/3$, $\eta =8/3$, and $\eta >8/3$ to produce physical solutions. Thus, while Eqs \eqref{critpt1} and \eqref{critpt2} must generally be solved numerically, we can use these constraints at the critical point to help identify the solutions of interest.

\begin{figure}
\center

\subfigure[]{
	\includegraphics [width=3.3in] {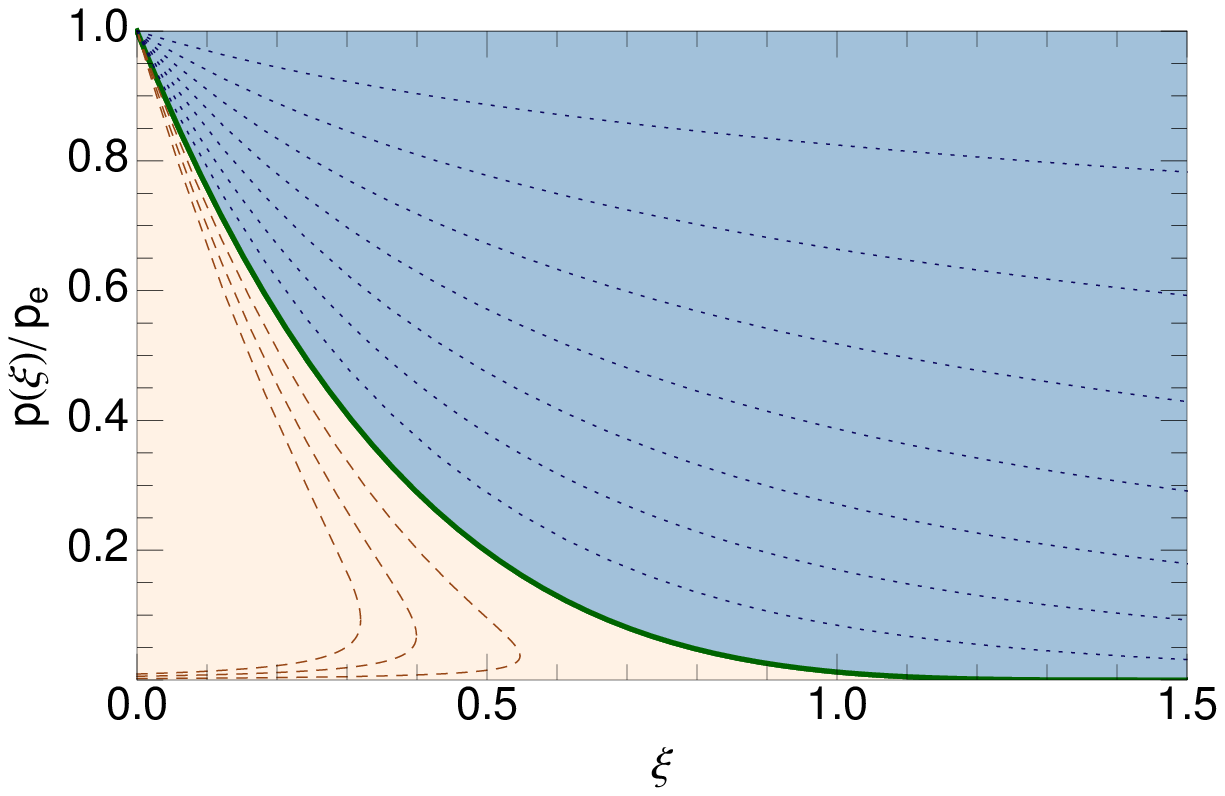}
	\label{fig:presscrash}
	}
\subfigure[]{
	\includegraphics [width=3.3in] {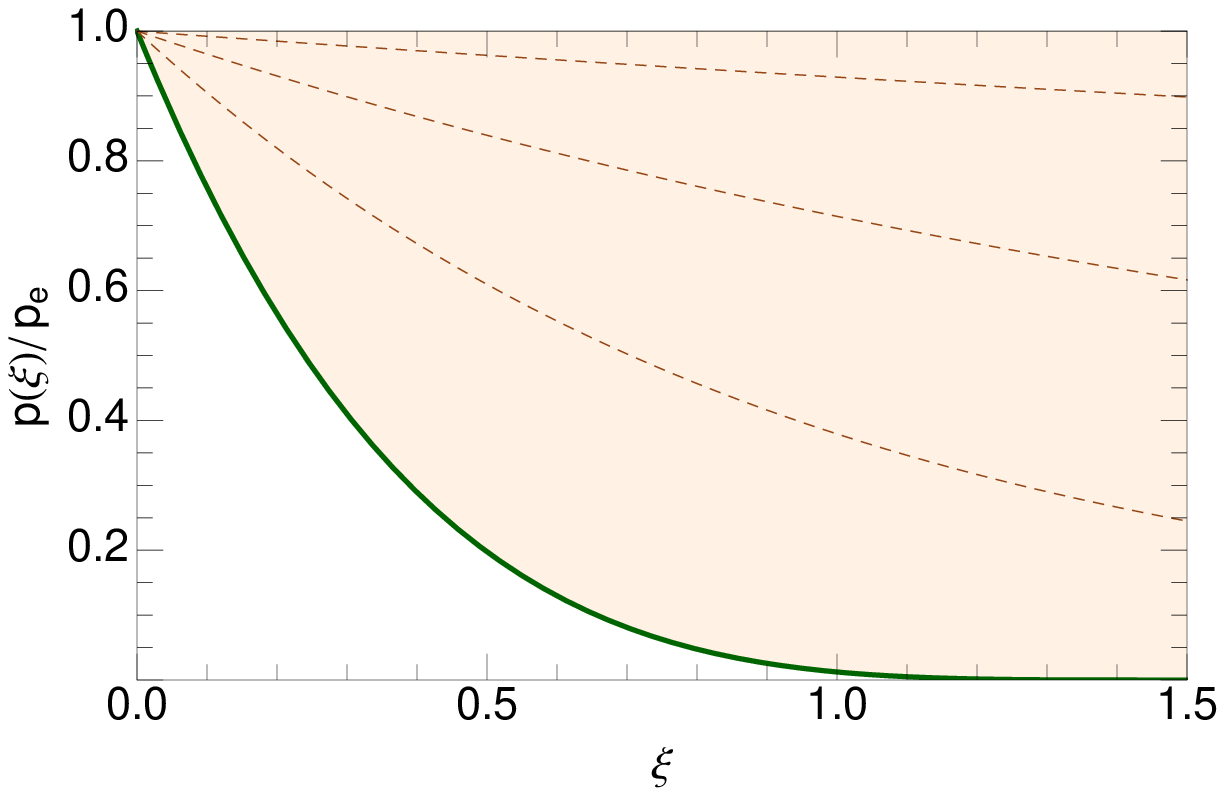}
	\label{fig:press}
	}
\caption{Pressure within the shock as scaled by the external pressure. The contact discontinuity is located at $\xi=0$. (a)~Curves for $2 < \eta < 4$ in increments of $1/6$ (beginning at $\eta=13/6$ and ending at $\eta=23/6$), for the special case that $h(\xi) = -\sqrt{2} g(\xi)$ everywhere. The value of $\eta$ increases when viewing the curves from lower left of the graph to the upper right. The curves for $\eta > 8/3$ are physical, while the curves for $\eta<8/3$ are not unless truncated. (b)~Physical solutions for $2 < \eta \le 8/3$ only, when the relation between $g$ and $h$ is allowed to vary.  The value of $\eta$ decreases when viewing the curves from lower left of the graph to the upper right. In both figures, the solid curve corresponds to $\eta=8/3$; the orange region contains curves for $\eta < 8/3$ and the blue region contains curves for $\eta > 8/3$.}
\label{fig:pressure}
\end{figure}

One such family of solutions occurs when $\mu = 1/2$ and $q(\xi)=w(\xi)$, i.e., when $h(\xi)=-\sqrt{2} g(\xi) $ everywhere. The solutions for $p \propto g(\xi)^4$ in this case are shown in Figure \ref{fig:presscrash}.

The most interesting result evident is that the structure of the solutions is divided based on the value of $\eta$. For $\eta = 8/3$ the solution is analytic and linear: $q=w= 1 - 2 \xi / 3$. This implies that the pressure decreases monotonically from the outer wall of the boundary layer ($\xi = 0$) inward, with $p \propto g(\xi)^4 \propto (1 - 2\xi/3)^4$, while the Lorentz factor $\Gamma$ increases linearly inward.

For $\eta > 8/3$, within this special family, the solutions decrease monotonically and asymptote to zero, with the steepest decrease in $q$ and $w$ occurring for $\eta \approx 8/3$. The curve for $\eta = 4$ is roughly constant at $q = w \approx 1$.

Solutions for $\eta < 8/3$ within this special family are also possible, but they are not single-valued: at the critical point the derivatives diverge and the solutions do not continue to larger values of $\xi$. For these solutions to be physical, the boundary layer would have to be truncated at a point before where the solutions crash and become double-valued.

By looking beyond this special family of solutions, it is possible to identify a solution for every $2<\eta<4$ wherein the pressure is a monotonically decreasing, single-valued function of $\xi$.

\begin{figure}
\begin{flushleft}
\begin{minipage}[]{0.55\linewidth}
	\begin{flushleft}
	\includegraphics [width=2.0in] {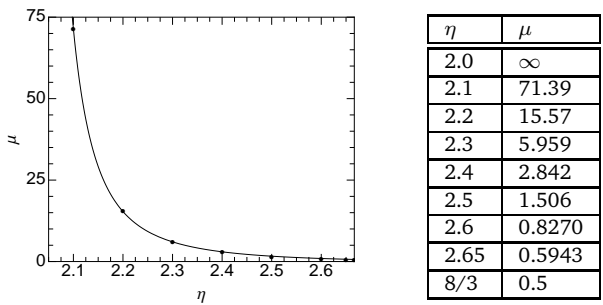}
	\end{flushleft}
\end{minipage}
\hspace{1cm}
\begin{minipage}[]{0.25\linewidth}
	    \begin{tabular}{| l | l |}
	    \hline
  	  $\eta$ & $\mu$ \\ \hline \hline
  	  2.0 & $\infty$ \\ \hline
  	  2.1 & 71.39 \\ \hline
  	  2.2 & 15.57 \\ \hline
  	  2.3 & 5.959 \\ \hline
 	  2.4 & 2.842 \\ \hline
 	  2.5 & 1.506 \\ \hline
  	  2.6 & 0.8270 \\ \hline
 	  2.65 & 0.5943 \\ \hline
 	  8/3 & 0.5 \\ \hline
 	   \end{tabular}
\end{minipage}
\end{flushleft}

\caption{Some values of $\mu$ as a function of $\eta$ for $2<\eta<8/3$, found empirically by seeking, for each value of $\eta$, the solution that passes through the critical point. A numerical fit to these data is shown on the left, and a table of values is given to the right.}
\label{fig:mu}
\end{figure}

\begin{figure}
\center
\includegraphics [width=3.3in] {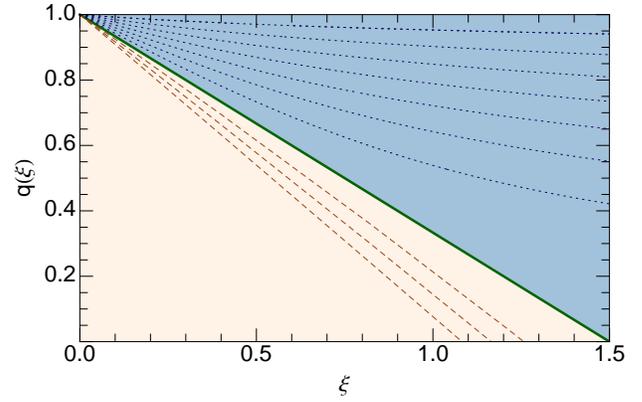}

\caption{Physical solutions for $q(\xi)$ within the shock, where the relation between $g$ and $h$ has been allowed to vary. The contact discontinuity is located at $\xi=0$. Curves for $2 < \eta < 4$, in increments of $1/6$ (again beginning at $\eta=13/6$ and ending at $\eta=23/6$), are plotted. The value of $\eta$ increases when viewing the curves from lower left of the graph to the upper right, with the solid curve corresponding to $\eta=8/3$. The coloring is the same as that in Figure \ref{fig:pressure}.}

\label{fig:q}
\end{figure}

In the case of $\eta \ge 8/3$, the unique physical solutions are the solutions that belong to the special family where $\mu=1/2$ and $q=w$. For $\eta<8/3$, however, the value of $\mu$ is not fixed. Instead, for each value of $\eta$ there is a single value of $\mu$ that yields the physical solution that traverses the critical point. For these solutions, $\mu$ varies from $\mu = 1/2$ for $\eta = 8/3$ to $\mu = \infty$ for $\eta=2$ (see Figure \ref{fig:mu}). When $\eta < 8/3$ the solutions for $w$ asymptote to zero, but the solutions for $q$ become negative at finite $\xi$, as in the $\eta = 8/3$ case (where $w$ and $q$ reach zero simultaneously). The steepest decrease in $q$ thus occurs for $\eta \approx 2$, whereas the steepest decrease in $w$ now occurs for $\eta \approx 8/3$. The physical pressure solutions for $\eta<8/3$ are plotted in Figure \ref{fig:press}, and the physical solutions for $q(\xi)$ (describing the spatial dependence of the transverse velocity) are plotted in Figure \ref{fig:q}.

Thus we find solutions that all display monotonically decreasing pressure from the contact discontinuity across the boundary layer, as we would expect for a pressure gradient arising from centripetal acceleration. Solutions for $\eta \approx 2$ have a nearly constant pressure across the boundary layer, but the pressure profiles become steeper as $\eta$ increases to $ 8/3$. The pressure for $\eta = 8/3$ is the only solution that goes to zero at a finite value, thereby forming a boundary layer of thickness $\xi \sim 1$, or $\Delta \theta \sim 1/\Gamma$. Above $\eta = 8/3$ the pressure profiles begin to decrease more gradually again, and for $\eta \approx 4$ the profiles again approach a constant.

Physically, this has very interesting implications. For $\eta = 8/3$, the pressure dropping to zero at a fixed value of $\xi$ implies that all of the jet material has piled up in the outer region of the boundary layer. Thus the jet is a hollow cone: the material is concentrated against the outer wall of the jet, and the inner region is essentially evacuated. This ``edge pileup'' has been studied as a possibility in quasar jets previously, e.g. \citealp{Zakamska08}. Observationally, Zakamska et al. suggested this phenomenon as an alternative explanation for observed edge-brightening in jets, which is more commonly interpreted as a result of the Kelvin-Helmholtz instability occurring at the jet boundary.

For values of $\eta$ near $8/3$, while the material is not immediately all piled up against the wall, the structure is not dissimilar. The region against the outer wall has the highest pressure, and the sharp pressure decrease interior to the outer wall suggests that most of the material is still in the outer region. While not truly a hollow cone, this structure could still account for observed edge-brightening in jets. Moving in either direction away from $\eta=8/3$, however, this region becomes broader and the jet becomes more evenly distributed.

For large and small enough values of $\eta$, the pressure gradient becomes small and the deviation of the pressure within the shock from the external pressure is minimal, suggesting that the Kompaneets approximation is a valid approximation in this region. For values of $\eta$ nearer to $8/3$, however, the Kompaneets approximation is clearly not a reasonable one.

The fact that the pressure profile begins level at $\eta=2$, steepens as $\eta$ approaches $8/3$, and then levels out again as $\eta$ approaches $4$ can be explained as the result of a tradeoff between two opposing effects: as $\eta$ increases, the degree of collimation decreases. This causes a decrease in the  centripetal force, and therefore the pressure gradient should decrease accordingly. But an increase in $\eta$ also means an increase in the rate of jet acceleration (as measured by the increase of $\Gamma$ with $r$), which will result in a steeper pressure gradient across the boundary. This acceleration effect appears to be stronger in the region where $2 < \eta < 8/3$, but the collimation effect wins out in the region where $8/3 < \eta < 4$. The two effects are in balance when $\eta= 8/3$,  leading to maximal compression of the gas against the wall of the jet.

This dividing behavior at $\eta = 8/3$ is not entirely unexpected, as can be shown by a quick calculation. The power carried by a pressure-dominated jet is given by
	\B
	L \propto p \Gamma^2 A,
	\E
where $p$ is the pressure and $A$ is the cross-sectional area of the region carrying most of the power.  If we assume that the power is concentrated in an outer layer of width $\Delta \theta = 1/\Gamma$, then viewed end-on the cross-sectional area of that ring would be $A \propto r^2/\Gamma$, supposing that the jet is conical to first order with very slow collimation. Using the external pressure profile and employing the relativistic Bernoulli equation, we see then that
	\B
	L \propto r^{-\eta/2} A = r^{-3\eta / 4 + 2}. \label{lumscale}
	\E
If we demand that the jet power be distance-independent, thus containing a finite amount of energy in the boundary layer, then this scaling implies that $\eta = 8/3$ must be true.

For $\eta<8/3$, we have $L$ scaling with radius to some positive power, such that for $r \rightarrow \infty$, $L \rightarrow \infty$. Thus, given a roughly conical jet, the power carried cannot remain constant with $\eta < 8/3$. Reexamining Eq \eqref{lumscale}, however, we can see that if the area scaled as some smaller power of $r$ rather than as described, then the jet power could be maintained as a constant despite the lower value of $\eta$. Thus, for $\eta<8/3$, the jet must become more strongly collimated, causing the cross-sectional area to grow more slowly than in the conical case, in order for the jet power not to diverge. 

For $\eta > 8/3$, it would appear that $L\rightarrow 0$ as $r \rightarrow \infty$. This is misleading, however, given that for this range in $\eta$, the integrals over $\xi$ of the energy contained within the boundary layer diverge. Thus, in a total spatial integral of $L$ over both $\xi$ and $r$, the $r$-scaling of $L \rightarrow 0$ and the $\xi$-scaling of $L \rightarrow \infty$ can combine to counteract each other, making it possible to obtain physical solutions under which a finite amount of energy is contained within the boundary layer after all.


\subsection{Extension to the Kompaneets Model}

This self-similar construction provides an important view of the behavior within the boundary layer, but to properly understand observations of the large-scale AGN jets that we seek to model, we must examine how the boundary-layer physics fits into that of the jet as a whole.

To this end, we now attempt to extend the Kompaneets model by repeating \S 2 with the addition of a pressure gradient across the boundary layer. The form of this added pressure gradient will come from our self-similar boundary-layer model, and in this section we work in cylindrical coordinates to facilitate the matching.

Applying the self-similar model to the structure developed in \S 2 inherently renders the problem no longer self-similar, as length scales are introduced into the problem. Nonetheless, the combination of methods is instructive in the description of general trends that are expected when adding a pressure gradient to the Kompaneets model.

Because $\xi$ is a function of distance from the contact discontinuity $(\theta_c - \theta)$, using the self-similar pressure solutions in the model in \S 2 requires that we already know the shape of this outer wall. As the self-similar model was developed with the assumption of an approximately conical contact discontinuity, we solve this problem by fitting a conical solution to each contact discontinuity found in \S 2 using the Kompaneets approximation and then using this fit as a fixed input for the location of the outer wall. As was shown in \S 2, approximating the contact discontinuity as conical is justified in the case of jets where the shock never closes to the axis, or in the region of the jet before the shock closes. It is this regime that we study.

\begin{figure}
\centering

\subfigure[$\eta=8/3, \theta_0=1/50, \Gamma_0=50$]{
	\begin{tabular}{cc}
		\includegraphics [width=1.5in]{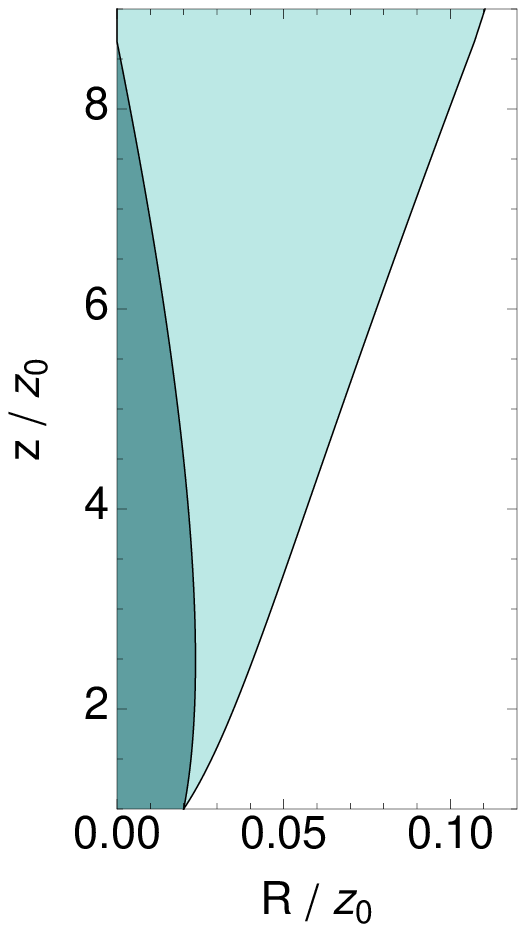} &
		\includegraphics[width=1.5in]{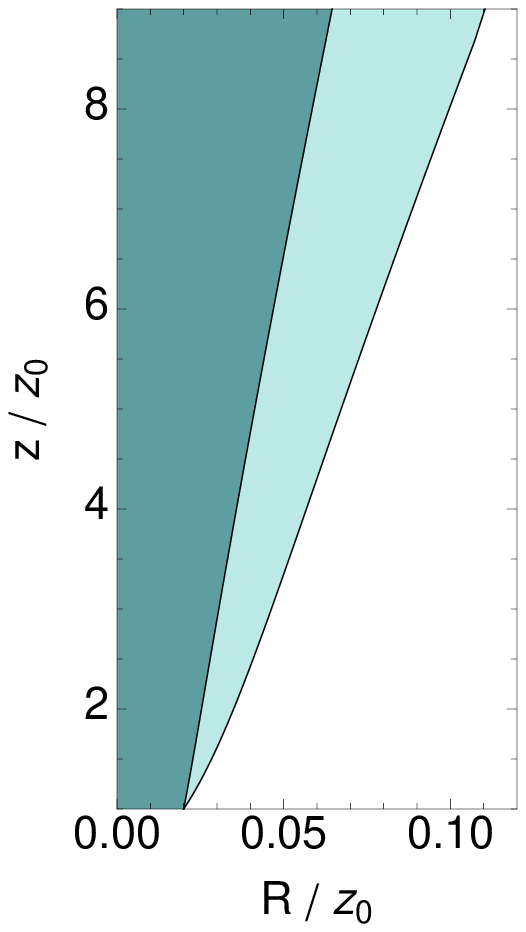}
	\label{fig:kludgeclose}
	\end{tabular}
}	\\
\subfigure[$\eta~=~8/3, \theta_0~=~2/10, \Gamma_0~=~10$]{
	\begin{tabular}{cc}
		\includegraphics [width=1.5in]{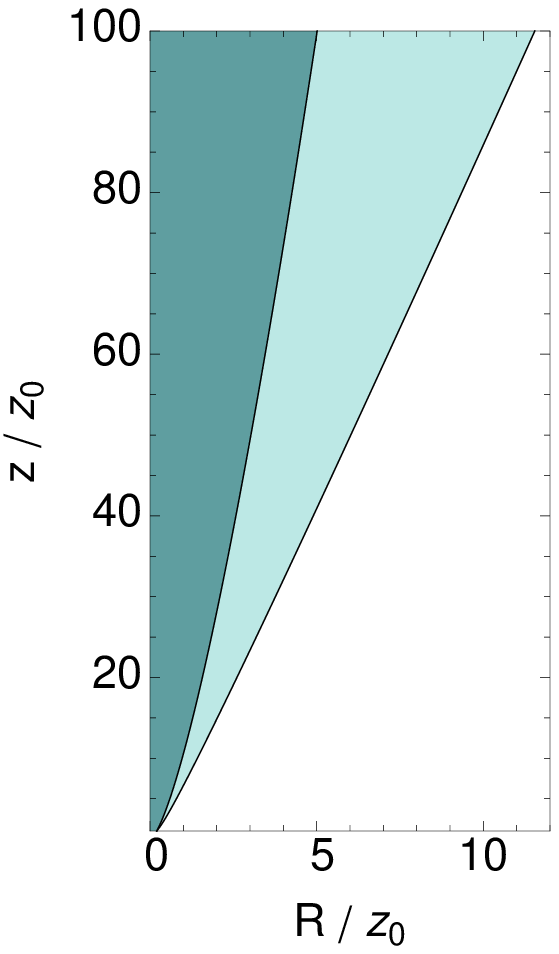} &
		\includegraphics [width=1.5in]{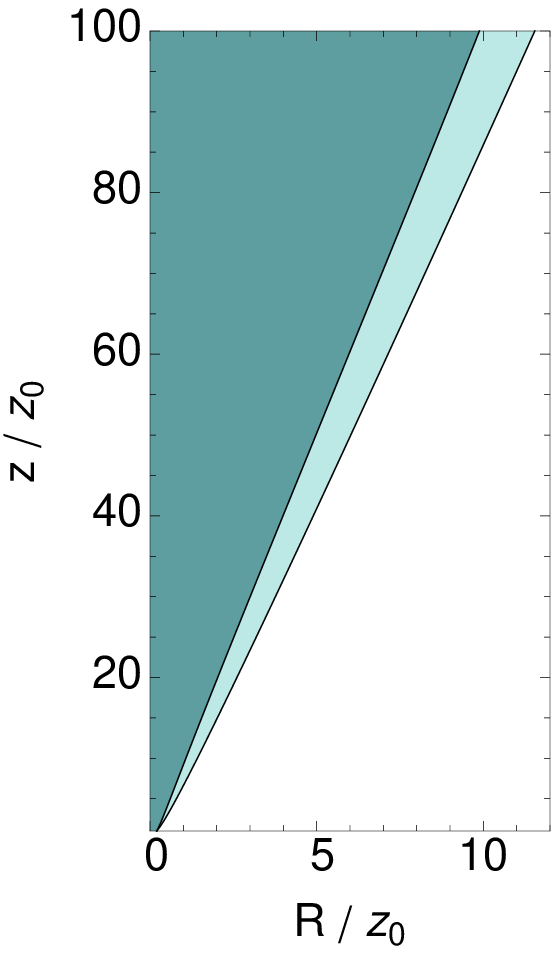}
	\label{fig:kludgeopen}
	\end{tabular}
}	

\caption{The shock front and contact discontinuity for two cases: one in which the jet originally closes (a) and one in which it originally remains open (b). Left plots demonstrate the jet under the Kompaneets approximation, from \S 2. Right plots indicate the new position of the shock front after the addition of the pressure gradient. The contact discontinuity remains unchanged in this approximation.}
\label{fig:kludge}
\end{figure}

We examine three representative cases --- one for $\eta<8/3$, one for $\eta=8/3$, and one for $\eta>8/3$ --- and use these to infer the general behavior of the jet shape after the addition of a pressure gradient. 

For $\eta=7/3$, the physical self-similar solution is that for which $\mu=4.575$. From this we fit an analytic function to the numerical solution for $g$, of the form
	\B
	g(\xi) &=& a + \frac{b}{(\xi + c)} + \frac{d}{(\xi+e)^2}, \label{anfit}
	\E
where $a, b, c, d,$ and $e$ are constant parameters determined by the fit. We then modify the expression for the pressure within the boundary layer so that $p_s \propto r^{-\eta}~g^4(\xi)$. We find that the results display the expected physical effect of adding this pressure gradient: the shock front moves outward to match the decreased pressure.

For $\eta=8/3$, the solution for $g(\xi)$ was analytic:
	\B
	g(\xi) &=& 1 - \dfrac{2}{3} \xi. \label{eta83g}
	\E
	
Using this function to add a pressure gradient into the case from \S 2, we again find that the shock wall moves outward to match the decreased pressure function. Two examples are shown in Figure \ref{fig:kludge}: one in which the shock front initially closed to the axis in the Kompaneets approximation, and one in which it initially remained open. Due to the approximation of a conical contact discontinuity breaking down beyond the point where the shock front initially closes, we cannot draw conclusions from these models in that regime. Nonetheless, it is clear that the addition of a pressure gradient either causes the shock front to converge to the axis at a distance further down the jet axis, or else it prevents the shock front from ever reaching the axis. Either way, one can see from Figure \ref{fig:kludge} that the addition of the pressure gradient causes the shocked boundary layer to become thinner.

For $\eta = 3$, we can again fit the function given by Eq \eqref{anfit} to the numerical solution of $g(\xi)$. This time the physical solution corresponds to $\mu=1/2$. Again modifying the Kompaneets solution of \S 2 with this pressure gradient, we find that here too the shock moves outward and forms a thin boundary layer. As with the $\eta = 8/3$ solutions, cases that originally closed to the axis in the Kompaneets approximation either close further down the axis or no longer close. 

We would hope that the energy within the shocked layer increases with radius, despite the message of the luminosity scaling worked out at the end of \S 3. As a test of this, for $\eta=3$, we can integrate the total energy contained in the boundary layer in a slice across the jet at a fixed $z$:
	\B
	L \propto \int_{R_j}^{R_c} p \Gamma^2 R dR.
	\E

\begin{figure}
\center
\includegraphics [width=3.3in] {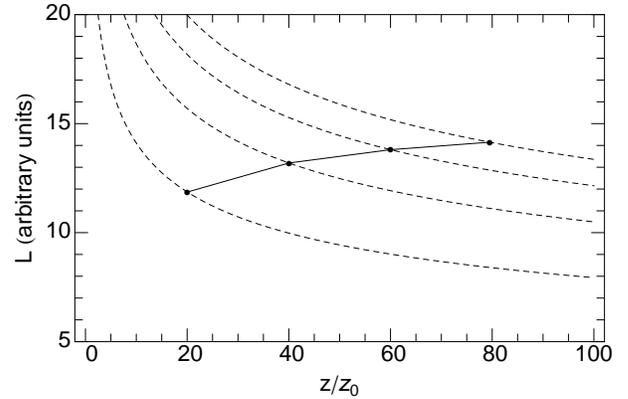}
\caption{The total integrated energy contained within the boundary layer at a given height $z$ for $\eta=3$. $L$ is given in arbitrary units. The dashed curves through the data points correspond to what the total-energy curve through each data point would be if $\xi$ were instead a fixed value along the shock.}
\label{fig:lcurves}
\end{figure}

Comparing this enclosed energy at a few different values of $z$ (see Figure \ref{fig:lcurves}), we can see that although the total energy within the layer would be decreasing if $\xi$ were fixed along the shock (the dashed curves show the fixed-$\xi$ energy through a given value of $z$), the net energy instead increases as we go to higher $z$ because we move to different values of $\xi$ in the process of traveling along the shock. Thus the solution is entirely physical: the energy behind the shock increases with an increase in distance from the source, as is required.\\
	
\section{Conclusion}

\indent We have evaluated the shape of the shocked boundary layer of a hot jet with an ultrarelativistic equation of state, in the case that it is injected into an ambient medium that has a power-law pressure profile of $p \propto r^{- \eta}$ with $2 < \eta < 4$. Using the shock jump conditions and momentum conservation and assuming that the pressure remains constant across the boundary layer, we found that whether the shocked layer closes to the axis --- and where it closes, if it does --- is dependent upon the values of the initial pressure ratio, the power-law index, the initial bulk Lorentz factor, and the product of the initial opening angle and initial Lorentz factor. 

We also found that, in the Kompaneets approximation, the contact discontinuity asymptotes to the shape $R_c \propto z^{\eta/4}$ in the case where the shock has closed to the axis, but to a conical shape, $R_c \propto z$, in the case of a non-closing shock or in the region of the jet before the shock has closed.

Due to the expectation of a pressure gradient arising from the centripetal force created by the slight collimation of the jet, we then created a self-similar model of the boundary layer that allows for a pressure gradient across the layer. From this we found solutions for which pressure decreases monotonically across the boundary layer for all values of the power-law index $\eta$.

An $\eta$ of 8/3 constitutes a special solution where the pressure goes to zero at a finite distance from the outer wall. For values of $\eta$ greater or less than 8/3, the pressure approaches zero asymptotically. The curve for pressure steepens  for values of $\eta$ near 8/3 and approaches a constant value for $\eta \rightarrow 2$ or $\eta \rightarrow 4$.

The drastic decrease of the pressure inwards from the contact discontinuity for values of $\eta$ near 8/3 suggests that most of the material is pushed up against the outer wall, creating a hollow-cone structure for the jet that may be observable as edge-brightening. It also indicates that the Kompaneets approximation is not valid for these values of $\eta$. For values of $\eta$ near 2 or 4, however, the Kompaneets approximation may be reasonable, since the pressure decrease is very gradual.

To better understand these results, we then revised the Kompaneets solutions to include the pressure gradient of the self-similar boundary-layer model. We held the outer wall fixed and solved for the new position of the shock front by pressure-matching across the shock jump. We found that the addition of the pressure gradient caused the shock front to move outward in order to match the lower pressure, resulting in a thinner boundary layer and preventing the shock from closing to the axis in places where it originally had done so. We also confirmed that the total energy contained within the boundary layer was an increasing function of height $z$, as is required.

The inherent difficulty of examining this problem analytically required us to make several simplifying assumptions over the course of this work. One such assumption, made in \S 3 to obtain the boundary-layer solutions, is that all the material within the boundary layer is on the same adiabat. This would be true if the majority of the material entered the boundary layer at roughly the same location, however this is not the case in general. But for the boundary-layer solutions where the pressure drops rapidly ($\eta$ near $8/3$), most of the material must be pushed up against the contact discontinuity early on. This suggests that it must all have entered the boundary layer near the jet base, which demonstrates self-consistency with the same-adiabat assumption for this regime.~Nonetheless, in future work we intend to explore the possibility of treating the jet material on different adiabats.

It is widely believed that magnetic effects contribute to the collimation of relativistic jets (see e.g. \citealp{KomissarovNumSim99}). While the models discussed here are purely hydrodynamic, they can have important applications both in interpreting the data obtained in three-dimensional general relativistic magnetohydrodynamics simulations that provide a self-consistent description of the jet launching mechanism (e.g. \citealp{Beckwith08a,Beckwith09,McKinney09}) and in improving the physical content of boundary conditions imposed in simulations that study the role of magnetic fields in large-scale jet collimation (see e.g. \citealp{KomBarkVla07,KomVlaKon09,Komissarov11}).

In the former case, the Poynting-flux dominated jet is sheathed by an unbound outflow (\citealp{Hawley06}), which may have important implications for the operation of current-driven instabilities (\citealp{McKinney09}). However, the numerical resolution used in these simulations is generally insufficient to adequately resolve the sharp density and pressure gradients present in these regions. Combining the models presented here with the conditions present in the ambient medium at the base of the jet (e.g. the disk corona) in these simulations will allow us to assess the extent to which numerical resolution affects simulation outcomes.

In the case of simulations that study the role of magnetic fields in large-scale jet collimation, the work presented here will allow a more complete physical treatment of the effects of the ambient medium. In these simulations (\citealp{KomBarkVla07,KomVlaKon09,Komissarov11}), this boundary is generally treated as a rigid wall, where fluid quantities (such as gas density and pressure) are simply copied across the boundary, while the normal component of vector quantities (e.g. velocity and magnetic field) is reflected. The models presented here will allow improvement of this treatment by specifying jump conditions for hydrodynamic quantities at this boundary, consistent with the shape of the wall --- thereby including the collimating effect of the ambient medium.

These calculations mark the first step towards a more complete treatment of both the action of the external medium and magnetic effects in collimating and accelerating relativistic jets. The next step in this work is to include the effect of a toroidal magnetic field, allowing for magnetic confinement. Together, this work and its magnetized extension will provide us with an equilibrium model from which we can explore instabilities and radiative mechanisms of the jet near its base, leading to improved constraints on jet dissipation and associated radiative signatures. Through this work we hope to further our understanding of jet collimation and, in particular, boundary-layer behavior.


\section*{Acknowledgements}

We thank Greg Salvesen, Sean O'Neill, and Jake Simon for valuable discussions, Fran\c{c}ois H\'ebert for numerical computation advice, and Krzysztof Nalewajko for comments on this manuscript. This work was supported in part by NSF grant AST-0907872 and NASA's Fermi Gamma-ray Space Telescope Guest Investigator program. \\

\newpage




\label{lastpage}

\end{document}